\documentclass[conference]{IEEEtran}
\IEEEoverridecommandlockouts
\usepackage{cite}
\usepackage{amsmath,amssymb,amsfonts}
\usepackage{graphicx}
\usepackage{textcomp}
\usepackage{xcolor}
\usepackage{graphicx}%
\usepackage{multirow}%
\usepackage{amsmath,amssymb,amsfonts}%
\usepackage{amsthm}%
\usepackage{mathrsfs}%
\usepackage{xcolor}%
\usepackage{textcomp}%
\usepackage{manyfoot}%
\usepackage{booktabs}%
\usepackage{algorithm}%
\usepackage{algpseudocode}%
\usepackage{listings}%
\usepackage{tabularx}%
\usepackage{colortbl}%
\usepackage{float}%
\usepackage{svg}%
\usepackage{array}%
\usepackage{url}%
\usepackage{times}%
\usepackage{comment}%
\usepackage{hyperref}

\def\BibTeX{{\rm B\kern-.05em{\sc i\kern-.025em b}\kern-.08em
    T\kern-.1667em\lower.7ex\hbox{E}\kern-.125emX}}

\usepackage{fancyhdr}
\begin{document}

\title{A Combined Channel Approach for Decoding Intracranial EEG Signals: Enhancing Accuracy through Spatial Information Integration\\}

\author{
\IEEEauthorblockN{1\textsuperscript{st} Maryam Ostadsharif Memar}
\IEEEauthorblockA{\textit{Department of Electrical and Computer Engineering} \\
\textit{Isfahan University of Technology} \\
Isfahan, Iran \\
m.ostadsharif@ec.iut.ac.ir}
\and
\IEEEauthorblockN{2\textsuperscript{nd} Navid Ziaei}
\IEEEauthorblockA{\textit{Department of Electrical and Computer Engineering} \\
\textit{Isfahan University of Technology} \\
Isfahan, Iran \\
n.ziaei@ec.iut.ac.ir}
\and
\IEEEauthorblockN{3\textsuperscript{rd} Behzad Nazari}
\IEEEauthorblockA{\textit{Department of Electrical and Computer Engineering} \\
\textit{Isfahan University of Technology} \\
Isfahan, Iran \\
nazari@iut.ac.ir}
}

%\maketitle
%--------------------------- Apply Header and Footer -------------------------------
% Note: You must manually add \thispagestyle{fancy} right after \maketitle 
% (just before the \begin{abstract} statement)
% to ensure the header and footer appear on the first page.
\maketitle
\thispagestyle{fancy}
%\begin{abstract}
% .....
%-----------------------------------------------------------------------------------

\begin{abstract}
Intracranial EEG (iEEG) recording, characterized by its high spatial and temporal resolution and superior signal-to-noise ratio (SNR), enables the development of precise brain-computer interface (BCI) systems for neural decoding. However, the invasive nature of the procedure significantly limits the availability of iEEG datasets, both in terms of the number of participants and the duration of recorded sessions. To overcome this, we propose a single participant machine learning model optimized for decoding iEEG signals. The model employs 18 key features and operates in two modes: best channel and combined channel. The combined channel mode integrates spatial information from multiple brain regions, resulting in superior classification performance. Evaluations across three datasets—Music Reconstruction, Audio Visual, and AJILE12—demonstrate that the model in combined channel mode consistently outperforms the best channel mode across all classifiers. In the best-performing cases, Random Forest achieved an F1 score of 0.81 ± 0.05 in the Music Reconstruction dataset, 0.82 ± 0.10 in the Audio Visual dataset, and XGBoost achieved an F1 score of 0.84 ± 0.08 in the AJILE12 dataset. Additionally, the analysis of brain region contributions in combined channel mode revealed that the model can identify relevant brain regions, aligned with physiological expectations, for each task and effectively combine the data from electrodes in these regions to achieve high performance. These findings underscore the resulting of integrating spatial information across brain regions to improve task decoding, offering new avenues for advancing BCI systems and neurotechnological applications.

\end{abstract}


\section{Introduction}
Brain-computer interfaces (BCIs) have emerged as a transformative technology, enabling direct communication between the brain and external devices. They hold immense potential for assisting individuals with neurological conditions such as amyotrophic lateral sclerosis (ALS), Parkinson's disease, and communication disorders like apraxia of speech \cite{ref39}. By decoding neural activity, BCIs can translate brain signals into control commands, facilitating communication and mobility for individuals with paralysis.

Neuroscientists use various techniques to record brain signals. Intracranial electroencephalography (iEEG) stands out among various neural recording techniques due to its high spatial and temporal resolution and superior signal-to-noise ratio (SNR). Unlike non-invasive methods like scalp EEG, iEEG captures precise neural activity directly from cortical and subcortical brain areas. However, iEEG has limitations: its invasive nature restricts the duration of recording sessions, and electrode implantation is targeted based on clinical needs, resulting in variable electrode coverage across participants \cite{ref40}. This variability complicates the development of generalized models for decoding neural signals.

Given the expanding potential of iEEG data, researchers have pursued various approaches to analyze these complex signals. Existing approaches to decoding iEEG signals have employed both traditional machine learning techniques and deep learning models. Traditional methods often rely on handcrafted features, which transform raw signals into informative representations for classification. Common features include time-domain metrics (e.g., amplitude, variance), frequency-domain attributes (e.g., power in specific bands like alpha and gamma), and time-frequency characteristics from wavelet or Fourier analyses \cite{ref37}. Additionally, connectivity measures (e.g., coherence, phase-locking \cite{ref35}) and nonlinear dynamics (e.g., entropy measures \cite{ref34}) capture interactions and complexity in the brain \cite{ref36}. These features are then applied in classifiers such as SVM \cite{ref5}, Bayesian Linear Discriminant Analysis \cite{ref9}, and K-Nearest Neighbors (KNN) \cite{ref10}, as well as newer methods like the Bayesian Time-Series classifier \cite{ref3} and the Latent Variable Gaussian Processes model \cite{ref11}. Although these methods demonstrate notable effectiveness, employing them is challenging because they require careful decisions regarding feature selection, extraction, and classifier design.

Deep learning approaches include convolutional neural networks (CNNs), such as EEGNet \cite{ref44}, which use CNNs to extract spatiotemporal features from neural data. Other studies have applied recurrent neural networks (RNNs), such as Long Short-Term Memory (LSTM) networks \cite{ref12} and Bidirectional RNNs \cite{ref7}, to capture temporal patterns in neural activity. Additionally, transformer-based models, such as BRAINBERT \cite{ref6} and IEEG-HCT \cite{ref13}, improve the analysis of complex brain activity patterns. These models have shown promise in capturing complex neural patterns but typically require large amounts of training data to perform effectively. The limited size of iEEG datasets, due to the invasiveness of data collection and participant-specific electrode coverage, makes it challenging to apply deep learning methods without overfitting \cite{ref45}.

To address these challenges, we propose a single-participant machine learning model optimized for decoding iEEG signals with limited data. 
Our approach involves extracting a comprehensive set of 18 features from both time and frequency domains, capturing essential characteristics of the neural signals. We then evaluate five classification algorithms—logistic regression, naive Bayes, random forest, SVM, and XGBoost—to determine the most effective methods for this task. A key innovation of our work is the introduction of two operational modes for the model:

\begin{enumerate}
    \item \textbf{Best Channel Mode:} This mode identifies and leverages the most discriminative single-channel\footnote{The terms ``channel" and ``electrode" are used interchangeably in this study, with ``channel" referring to the signals recorded from each implanted electrode.} data for each participant. While this approach is computationally efficient and straightforward, it may not fully leverage the spatial information available in iEEG recordings.

    \item \textbf{Combined Channel Mode:} This mode integrates information from multiple channels, utilizing neural data from different brain regions as spatial features to enhance classification performance. By combining data across channels, this mode addresses the challenge of variability in electrode placement across participant and maximizes the utilization of available neural information.
\end{enumerate}

 By examining the contributions of individual brain regions in the combined channel mode, we can gain insights into the neural mechanisms driving this performance. This analysis not only enhances model performance but also contributes to our understanding of brain function, potentially informing the development of more effective BCIs.

 We validate our approach using three publicly available iEEG datasets. Our results demonstrate that the combined channel mode consistently outperforms the best channel mode across all classifiers and datasets, highlighting the importance of integrating spatial information. Furthermore, our analysis reveals that certain brain regions, such as the superior temporal lobe in auditory tasks and the precentral gyrus in motor tasks, play critical roles in task decoding.

The subsequent sections of this paper are organized as follows: Section 2 outlines the feature extraction process and classifiers used in this model. We also present the training and evaluation methods for the model. In Section 3, we first introduce the public iEEG datasets used in this study and assess the model's performance on these datasets. Moreover, we explore the contribution of brain regions in the combined channel mode to understand neural patterns. Finally, in Section 4, we provide the conclusion.

\section{Material and Methods}
\subsection{Feature Extraction}
In this study, the gamma band (65–120 Hz) of the signal was initially extracted, and its Hilbert transform was computed. Subsequently, 18 features were derived from the processed data of each channel, comprising 12 time-domain features and 6 frequency-domain features. The selection of these features was based on their proven effectiveness in previous neural signal classification studies \cite{ref37, ref38} and their ability to represent key signal properties relevant to our tasks. A detailed description of the time and frequency domain features is provided in Tables \ref{tab1} and \ref{tab2}, respectively.

\begin{table}[t]
\centering
\caption{Time Domain Features}
\renewcommand{\arraystretch}{1.3}
\begin{tabular}{p{0.4\columnwidth} p{0.5\columnwidth}}
\hline
\textbf{Features} & \textbf{Description}  \\
\hline
1. Average \cite{ref15} & Averaging signal after smoothing \\
2. RMS \cite{ref14} & Calculating signal power \\
3. Max peak \cite{ref15} & Determining the maximum peak value \\
4. Variance \cite{ref15} & Computing the dispersion of data \\
5. Skewness \cite{ref14} & Assessing the asymmetry of data \\
6. Kurtosis \cite{ref14} & Describing the distribution shape \\
7. Autocorrelation \cite{ref14} & Correlation across time delays \\
8. Nonlinear energy \cite{ref16} & Measuring absolute value averaging \\
9. Spikes \cite{ref15} & Identifying sharp increases or peaks \\
10. HFD \cite{ref19} & Assessing irregularity or self-similarity \\
11. Shannon entropy \cite{ref17} & Measuring uncertainty and information content \\
12. Renyi entropy \cite{ref17} & Generalizing Shannon entropy to control sensitivity to probabilities \\
\hline
\end{tabular}
\label{tab1}
\end{table}

\begin{table}[t]
\centering
\caption{Frequency Domain Features}
\renewcommand{\arraystretch}{1.3}
\begin{tabular}{p{0.4\columnwidth} p{0.5\columnwidth}}
\hline
\textbf{Features} & \textbf{Description}  \\
\hline
1. Coastline \cite{ref15} & Summing absolute values of data point derivatives \\
2. Band powers \cite{ref14} & Finding the mean of power spectral density \\
3. Spectral edge frequency \cite{ref14} & Frequency at which 90\% of cumulative power is reached \\
4. Hjorth mobility \cite{ref16} & Quantifying the frequency dispersion of the signal \\
5. Hjorth complexity \cite{ref16} & Measuring the complexity of the signal \\
6. Spectral entropy \cite{ref18} & Quantifying randomness in frequency components \\
\hline
\end{tabular}
\label{tab2}
\end{table}

\subsection{Classification Models}
Following the feature extraction, we utilized five well-established machine learning classifiers to predict the corresponding labels: Logistic Regression \cite{ref27}, Naive Bayes \cite{ref28}, Random Forest \cite{ref27}, Support Vector Machine (SVM) \cite{ref26}, and XGBoost \cite{ref30}. These classifiers were chosen for their diverse approaches and demonstrated success in classification tasks. Ensemble methods such as Random Forest and XGBoost are effective for handling high-dimensional data by capturing complex feature interactions and reducing the risk of overfitting. Logistic Regression is well-suited for linearly separable data due to its simplicity and interpretability, while Support Vector Machines (SVM) are particularly effective for data with non-linear decision boundaries. Naive Bayes is a robust baseline classifier, especially effective for smaller datasets where the assumption of feature independence holds \cite{ref46}.

\subsection{Ensemble Method}
\label{ensemble method}
Each participant has a large number of channels, and using data from all channels to train a single classifier can result in overfitting due to the high dimensionality of the data. To address this, we adopted an ensemble approach, with a separate weak classifier trained on the features extracted from each individual channel. These single-channel classifiers can then be combined into a multi-channel classifier using the majority voting technique\cite{ref31}. Due to the exponential number of possible channel combinations, we implemented a greedy forward selection strategy instead of evaluating all subsets. This approach is computationally efficient and has been demonstrated to yield optimal solutions in ensemble construction \cite{ref41}. The process began by selecting the channel with the best performance. Then, we iteratively evaluated which additional channels, when added, could further improve performance. All possible combinations were assessed, and the one with the highest performance was selected. 

\begin{figure*}[t]
    \centering
    \includegraphics[width=\textwidth]{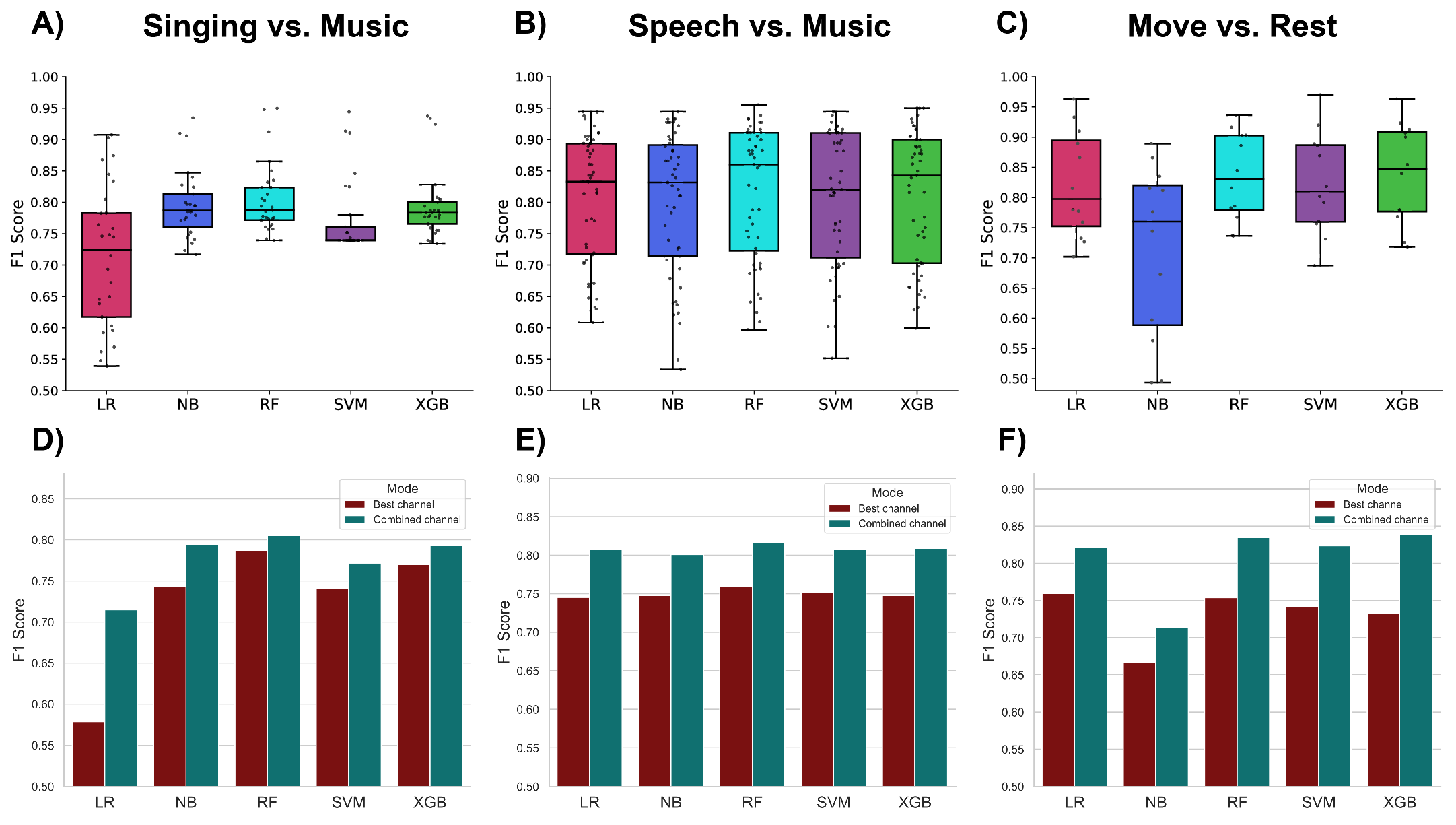}
    \caption{Performance comparison of the single-participant model in the best channel and combined channel modes across all classifiers and datasets. The classifiers used include Logistic Regression (LR), Naive Bayes (NB), Random Forest (RF), Support Vector Machine (SVM), and XGBoost (XGB). \textbf{(A, B, C)} Performance of the model in the combined channel mode using different classifiers. \textbf{(D, E, F)} Comparison of the model's performance between the best channel and combined channel modes. These results validate the importance of leveraging spatial information from all available channels (combined channel mode) compared to using only the best-performing channel (best channel mode).}
    \label{result}
\end{figure*}

\subsection{Model Training and Evaluation}
\label{model training}
To assess the performance of the single-participant model, we utilized 5-fold cross-validation. In each fold, the data from each electrode was split into a 64\%, 16\%, and 20\% ratio for the training set, validation set, and test set, respectively. After training each weak classifier (for the imbalanced dataset, we first balanced the training set using the Synthetic Minority Over-sampling Technique (SMOTE)\cite{ref32} before training), we evaluated the single-participant models in two modes: best channel and combined channel.

\textbf{Best Channel mode:} In this mode, we assessed the performance of each weak classifier using the validation sets and identified the best weak classifier based on its performance. Consequently, for each participant, we determined the channel that provided the most useful data for classification and considered the classifier of the best channel as the participant's model.

\textbf{Combined Channel mode:} To improve the performance of the single-participant model, it is essential to utilize data from all channels. In this mode, for each participant, we identified the effective set of channels using the validation set, as explained in Section \ref{ensemble method}. The outputs from the classifiers of these channels were then combined using the Majority Voting technique\cite{ref31} to achieve high performance.

\subsection{Evaluation Metrics}
To evaluate the performance of our model, we used three standard metrics, i.e., precision, recall, and F1 score:

\begin{equation}
\text{Precision} = \frac{TP}{TP + FP}
\end{equation}

\begin{equation}
\text{Recall} = \frac{TP}{TP + FN}
\end{equation}

\begin{equation}
\text{F1 Score} = 2 \cdot \frac{\text{Precision} \cdot \text{Recall}}{\text{Precision} + \text{Recall}}
\end{equation}

where \(TP\), \(FP\), and \(FN\) denote the true positives, false positives, and false negatives, respectively. These metrics are essential for providing a comprehensive assessment of model performance, particularly in the context of imbalanced datasets where traditional accuracy measures may be insufficient\cite{ref33}.

\section{Results}
\begin{figure*}[t]
    \centering
    \includegraphics[width=\textwidth]{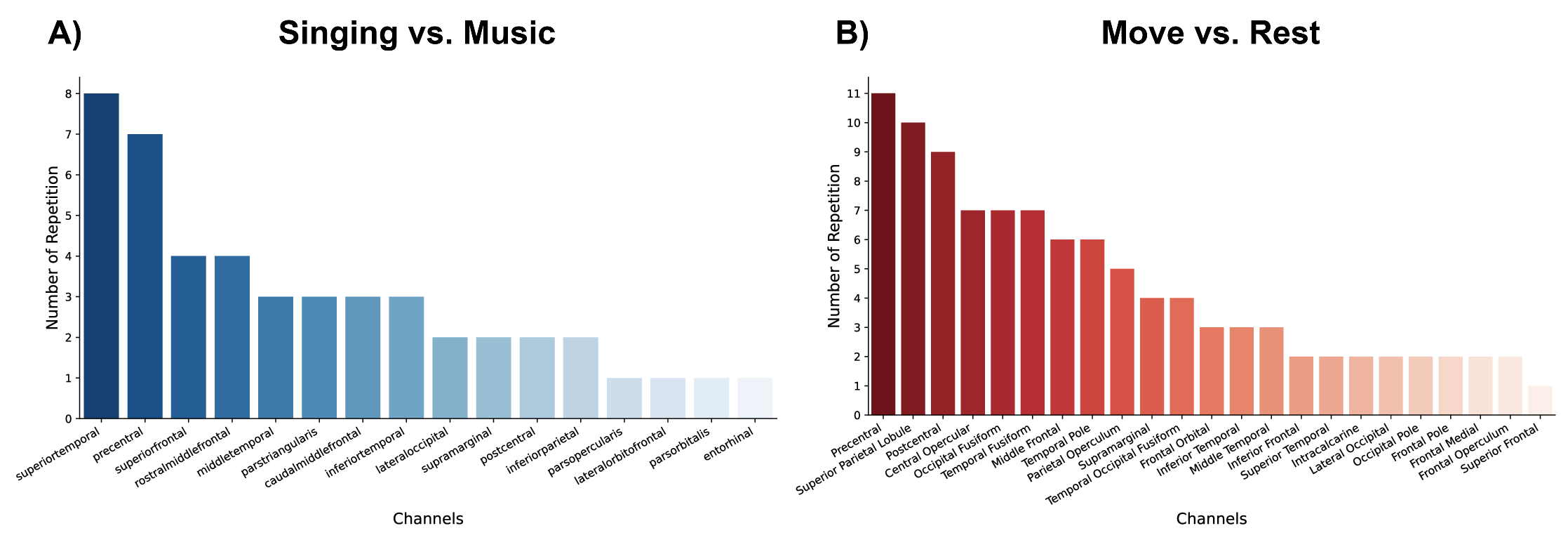}
    \caption{Brain regions contributing to the Singing vs. Music and Move vs. Rest tasks in the combined channel mode.}
    \label{brain region}
\end{figure*}

\subsection{Dataset Overview}
\label{dataset}
In this study, we evaluated the performance of the single-participant model using three datasets: the Music Reconstruction dataset \cite{ref21}, the Audio Visual dataset \cite{ref22}, and AJILE12 dataset \cite{ref23}.

\textbf{Music Reconstruction Dataset:} This dataset, described in \cite{ref21}, includes iEEG recordings from 29 neurosurgical participants who listened to rock music. Electrode placement was guided by clinical needs and positioned in either the right (11 participants) or left (18 participants) hemisphere. The musical stimuli comprised 32 seconds of vocal music and 2 minutes and 26 seconds of instrumental music. For analysis, we defined a classification task, Singing vs. Music, and segmented the recordings into 2-second trials. This resulted in 16 Singing trials and 73 Music trials per participant. The 2-second window duration was empirically chosen to optimize the model's performance.

\textbf{Audio Visual Dataset:} As described in \cite{ref22}, this dataset consists of iEEG recordings from 51 participants who watched a 6.5-minute French film. The film alternates between 30-second segments of actors' speech and music. In this dataset, we defined the classification task as Speech vs. Music. To achieve this, data were segmented into 2-second events, yielding 105 Music events and 90 Speech events per participant. The time window durations were also selected empirically.

\textbf{AJILE12 dataset:} This dataset, introduced in \cite{ref23}, comprises iEEG recordings from 12 participants undergoing clinical epilepsy monitoring, with simultaneous video recordings used to identify moments of upper-limb movement. Electrode placement was restricted to one hemisphere per participant, with 5 participants having electrodes on the right hemisphere and 7 on the left. The classification task for this dataset, Move vs. Rest, was defined based on the detection of upper-limb movements. Data were segmented into 2-second time windows centered around each trial, resulting in a minimum of 150 Move trials and 150 Rest trials per participant.

\subsection{Performance Analysis for Individual Participants}
We evaluated the performance of the single-participant model in two operational modes: best channel and combined channel, using the three datasets introduced in Section \ref{dataset}. Fig. \ref{result}(D, E, F) illustrates the comparative performance of the model in the best channel and combined channel modes across all classifiers for each dataset. As shown, the combined channel mode consistently outperformed the best channel mode. This outcome is consistent with expectations because the combined channel mode leverages the spatial information provided by multiple channels. By aggregating outputs from classifiers trained on data from various brain regions, the combined channel mode effectively captures a more comprehensive representation of the underlying neural activity, resulting in higher classification accuracy. 

We further evaluated the performance of the model in the combined channel mode using various types of classifiers, as depicted in Fig. \ref{result}(A, B, C). As shown in the figure, for the Singing vs. Music task, Random Forest achieved the highest performance (0.81 ± 0.05), outperforming Logistic Regression (0.71 ± 0.11) and SVM (0.77 ± 0.06), while slightly surpassing both Naive Bayes and XGBoost (0.79 ± 0.05). In the Speech vs. Music task, performance across classifiers was relatively uniform, with Logistic Regression, SVM, and XGBoost achieving 0.81 ± 0.1. Random Forest marginally outperformed them at 0.82 ± 0.1, and Naive Bayes had the lowest score at 0.80 ± 0.11. For the Move vs. Rest task, XGBoost exhibited the highest performance (0.84 ± 0.08), followed by Random Forest (0.83 ± 0.07). SVM and Logistic Regression both scored 0.82 ± 0.08, while Naive Bayes showed the lowest performance (0.71 ± 0.14)

\subsection{Brain Region Contributions in Combined Channel Mode}
To gain insight into neural patterns, we analyzed the contributions of different brain regions in the combined channel mode. For each participant, we identified the effective channels contributing to the majority voting process and mapped these channels to their corresponding brain regions. Fig.\ref{brain region} presents histograms illustrating the distribution of brain regions contributing to the combined channel models for each dataset. In this figure, we analyzed the repetition rate of brain regions included in the effective channel sets across participants to identify relevant brain regions for each classification task. This analysis was performed exclusively for the Music Reconstruction and AJILE12 datasets, as electrode position data is unavailable for the Audio-Visual dataset.

As shown in Fig. \ref{brain region}(A), the superior temporal lobe emerges as the most frequently recurring brain region across participants in the combined channel mode for the Singing vs. Music task. This finding suggests its central role in the neural dynamics relevant to the classification task. This result was expected, as the Singing vs. Music task is closely related to auditory processing, and this lobe plays an essential role in both auditory processing and language comprehension \cite{ref42}. Furthermore, this finding is consistent with prior research conducted on this dataset \cite{ref24}, \cite{ref21} as well as studies utilizing similar dataset \cite{ref7}.

As shown in Fig. \ref{brain region}(B), the precentral gyrus is the most prominent brain region across participants in the combined channel mode for the Move vs. Rest task, reflecting its key role in motor planning and execution. This is consistent with the established role of the precentral gyrus, which houses the primary motor cortex, as it is critically involved in movement tasks \cite{ref43}. Moreover, this finding aligns with other research using this dataset \cite{ref25}.

\section{Conclusion}
In this study, we developed a robust single-participant machine learning model for decoding intracranial EEG recordings. The model extracted 18 key features and employed classifiers such as SVM, logistic regression, Naive Bayes, random forest, and XGBoost. By leveraging spatial information from multiple brain regions, the model addresses the challenge of limited data availability in iEEG recordings. Through the evaluation of three distinct datasets, we consistently observed high performance, confirming the method's effectiveness across all datasets. Additionally, we demonstrated that the model can identify key brain regions crucial for decoding specific tasks, and that combining data from channels associated with these regions significantly enhances performance. These findings underscore the importance of targeted spatial feature integration in iEEG classification tasks, paving the way for more accurate and interpretable brain-computer interface applications.

\section*{Code and data availability}
The single participant model code utilized in this study is publicly available on \url{https://github.com/Navid-Ziaei/combined-channel-iEEG-decoder.git}. The code is designed to be used in conjunction with the following publicly accessible datasets: Audio Visual dataset (\url{https://openneuro.org/datasets/ds003688}), Music Reconstruction dataset (\url{https://zenodo.org/records/7876019}), and AJILE12 dataset (\url{https://figshare.com/projects/Generalized_neural_decoders_for_transfer_learning_across_participants_and_recording_modalities/90287}). Together, these resources enable the full reproduction of the main findings and figures presented in this study.

\section*{AI Usage Statement}
The authors declare that they used artificial intelligence tools only to improve and edit the text of the article.

% Bibliography
\bibliographystyle{IEEEtran}  % IEEE citation style
\bibliography{main_ref}       

\begin{thebibliography}{10}
\providecommand{\url}[1]{#1}
\csname url@samestyle\endcsname
\providecommand{\newblock}{\relax}
\providecommand{\bibinfo}[2]{#2}
\providecommand{\BIBentrySTDinterwordspacing}{\spaceskip=0pt\relax}
\providecommand{\BIBentryALTinterwordstretchfactor}{4}
\providecommand{\BIBentryALTinterwordspacing}{\spaceskip=\fontdimen2\font plus
\BIBentryALTinterwordstretchfactor\fontdimen3\font minus \fontdimen4\font\relax}
\providecommand{\BIBforeignlanguage}[2]{{%
\expandafter\ifx\csname l@#1\endcsname\relax
\typeout{** WARNING: IEEEtran.bst: No hyphenation pattern has been}%
\typeout{** loaded for the language `#1'. Using the pattern for}%
\typeout{** the default language instead.}%
\else
\language=\csname l@#1\endcsname
\fi
#2}}
\providecommand{\BIBdecl}{\relax}
\BIBdecl

\bibitem{ref39}
E.~C. Leuthardt, G.~Schalk, J.~R. Wolpaw, J.~G. Ojemann, and D.~W. Moran, ``A brain--computer interface using electrocorticographic signals in humans,'' \emph{Journal of neural engineering}, vol.~1, no.~2, p.~63, 2004.

\bibitem{ref40}
J.~Parvizi and S.~Kastner, ``Human intracranial eeg: promises and limitations,'' \emph{Nature neuroscience}, vol.~21, no.~4, p. 474, 2018.

\bibitem{ref37}
A.~K. Singh and S.~Krishnan, ``Trends in eeg signal feature extraction applications,'' \emph{Frontiers in Artificial Intelligence}, vol.~5, p. 1072801, 2023.

\bibitem{ref35}
R.~W. Thatcher, ``Coherence, phase differences, phase shift, and phase lock in eeg/erp analyses,'' \emph{Developmental neuropsychology}, vol.~37, no.~6, pp. 476--496, 2012.

\bibitem{ref34}
P.~Patel and R.~N. Annavarapu, ``Eeg-based human emotion recognition using entropy as a feature extraction measure,'' \emph{Brain informatics}, vol.~8, no.~1, p.~20, 2021.

\bibitem{ref36}
L.~Hu and Z.~Zhang, ``Eeg signal processing and feature extraction,'' 2019.

\bibitem{ref5}
H.~Liu, Y.~Agam, J.~R. Madsen, and G.~Kreiman, ``Timing, timing, timing: fast decoding of object information from intracranial field potentials in human visual cortex,'' \emph{Neuron}, vol.~62, no.~2, pp. 281--290, 2009.

\bibitem{ref9}
H.~Chang and J.~Yang, ``Genetic-based feature selection for efficient motion imaging of a brain--computer interface framework,'' \emph{Journal of Neural Engineering}, vol.~15, no.~5, p. 056020, 2018.

\bibitem{ref10}
S.~Paul, I.~Zabir, T.~Sarker, S.~A. Fattah, and C.~Shahnaz, ``Higher order statistics of bispectrum and mrp of ecog signals for motor imagery tasks classification,'' in \emph{2017 IEEE Region 10 Symposium (TENSYMP)}.\hskip 1em plus 0.5em minus 0.4em\relax IEEE, 2017, pp. 1--4.

\bibitem{ref3}
N.~Ziaei, R.~Saadatifard, A.~Yousefi, B.~Nazari, S.~S. Cash, and A.~C. Paulk, ``Bayesian time-series classifier for decoding simple visual stimuli from intracranial neural activity,'' in \emph{International Conference on Brain Informatics}.\hskip 1em plus 0.5em minus 0.4em\relax Springer, 2023, pp. 227--238.

\bibitem{ref11}
N.~Ziaei, J.~J. Stim, M.~D. Goodman-Keiser, S.~Sponheim, A.~S. Widge, S.~Krikorian, and A.~Yousefi, ``Latent variable double gaussian process model for decoding complex neural data,'' \emph{arXiv preprint arXiv:2405.05424}, 2024.

\bibitem{ref44}
V.~J. Lawhern, A.~J. Solon, N.~R. Waytowich, S.~M. Gordon, C.~P. Hung, and B.~J. Lance, ``Eegnet: a compact convolutional neural network for eeg-based brain--computer interfaces,'' \emph{Journal of neural engineering}, vol.~15, no.~5, p. 056013, 2018.

\bibitem{ref12}
M.~Rashid, M.~Islam, N.~Sulaiman, B.~S. Bari, R.~K. Saha, and M.~J. Hasan, ``Electrocorticography based motor imagery movements classification using long short-term memory (lstm) based on deep learning approach,'' \emph{SN Applied Sciences}, vol.~2, pp. 1--7, 2020.

\bibitem{ref7}
S.~L. Metzger, K.~T. Littlejohn, A.~B. Silva, D.~A. Moses, M.~P. Seaton, R.~Wang, M.~E. Dougherty, J.~R. Liu, P.~Wu, M.~A. Berger \emph{et~al.}, ``A high-performance neuroprosthesis for speech decoding and avatar control,'' \emph{Nature}, vol. 620, no. 7976, pp. 1037--1046, 2023.

\bibitem{ref6}
C.~Wang, V.~Subramaniam, A.~U. Yaari, G.~Kreiman, B.~Katz, I.~Cases, and A.~Barbu, ``Brainbert: Self-supervised representation learning for intracranial recordings,'' \emph{arXiv preprint arXiv:2302.14367}, 2023.

\bibitem{ref13}
M.~Yang, K.~Wang, C.~Li, R.~Qian, and X.~Chen, ``Ieeg-hct: A hierarchical cnn-transformer combined network for intracranial eeg signal identification,'' \emph{IEEE Sensors Letters}, 2024.

\bibitem{ref45}
C.~Brunner, N.~Birbaumer, B.~Blankertz, C.~Guger, A.~K{\"u}bler, D.~Mattia, J.~d.~R. Mill{\'a}n, F.~Miralles, A.~Nijholt, E.~Opisso \emph{et~al.}, ``Bnci horizon 2020: towards a roadmap for the bci community,'' \emph{Brain-computer interfaces}, vol.~2, no.~1, pp. 1--10, 2015.

\bibitem{ref38}
I.~Stancin, M.~Cifrek, and A.~Jovic, ``A review of eeg signal features and their application in driver drowsiness detection systems,'' \emph{Sensors}, vol.~21, no.~11, p. 3786, 2021.

\bibitem{ref15}
A.~Dalton, S.~Patel, A.~R. Chowdhury, M.~Welsh, T.~Pang, S.~Schachter, G.~OLaighin, and P.~Bonato, ``Development of a body sensor network to detect motor patterns of epileptic seizures,'' \emph{IEEE transactions on biomedical engineering}, vol.~59, no.~11, pp. 3204--3211, 2012.

\bibitem{ref14}
F.~Mormann, R.~G. Andrzejak, C.~E. Elger, and K.~Lehnertz, ``Seizure prediction: the long and winding road,'' \emph{Brain}, vol. 130, no.~2, pp. 314--333, 2007.

\bibitem{ref16}
S.~Mukhopadhyay and G.~Ray, ``A new interpretation of nonlinear energy operator and its efficacy in spike detection,'' \emph{IEEE Transactions on biomedical engineering}, vol.~45, no.~2, pp. 180--187, 1998.

\bibitem{ref19}
A.~H. Al-Nuaimi, E.~Jammeh, L.~Sun, and E.~Ifeachor, ``Higuchi fractal dimension of the electroencephalogram as a biomarker for early detection of alzheimer's disease,'' in \emph{2017 39th Annual International Conference of the IEEE Engineering in Medicine and Biology Society (EMBC)}.\hskip 1em plus 0.5em minus 0.4em\relax IEEE, 2017, pp. 2320--2324.

\bibitem{ref17}
A.~Feltane, G.~B. Bartels, J.~Gaitanis, Y.~Boudria, and W.~Besio, ``Human seizure detection using quadratic r{\'e}nyi entropy,'' in \emph{2013 6th International IEEE/EMBS Conference on Neural Engineering (NER)}.\hskip 1em plus 0.5em minus 0.4em\relax IEEE, 2013, pp. 815--818.

\bibitem{ref18}
H.~Viertio-Oja, V.~Maja, M.~Sarkela, P.~Talja, N.~Tenkanen, H.~Tolvanen-Laakso, M.~Paloheimo, A.~Vakkuri, A.~Yli-Hankala, and P.~Merilainen, ``Description of the entropytm algorithm as applied in the datex-ohmeda s/5 tm entropy module,'' \emph{Acta anaesthesiologica scandinavica}, vol.~48, no.~2, pp. 154--161, 2004.

\bibitem{ref27}
G.~James, ``An introduction to statistical learning,'' 2013.

\bibitem{ref28}
I.~Rish \emph{et~al.}, ``An empirical study of the naive bayes classifier,'' in \emph{IJCAI 2001 workshop on empirical methods in artificial intelligence}, vol.~3, no.~22.\hskip 1em plus 0.5em minus 0.4em\relax Citeseer, 2001, pp. 41--46.

\bibitem{ref26}
A.~J. Smola and B.~Sch{\"o}lkopf, ``A tutorial on support vector regression,'' \emph{Statistics and computing}, vol.~14, pp. 199--222, 2004.

\bibitem{ref30}
T.~Chen and C.~Guestrin, ``Xgboost: A scalable tree boosting system,'' in \emph{Proceedings of the 22nd acm sigkdd international conference on knowledge discovery and data mining}, 2016, pp. 785--794.

\bibitem{ref46}
M.-P. Hosseini, A.~Hosseini, and K.~Ahi, ``A review on machine learning for eeg signal processing in bioengineering,'' \emph{IEEE reviews in biomedical engineering}, vol.~14, pp. 204--218, 2020.

\bibitem{ref31}
F.~Leon, S.-A. Floria, and C.~B{\u{a}}dic{\u{a}}, ``Evaluating the effect of voting methods on ensemble-based classification,'' in \emph{2017 IEEE international conference on INnovations in intelligent Systems and applications (INISTA)}.\hskip 1em plus 0.5em minus 0.4em\relax IEEE, 2017, pp. 1--6.

\bibitem{ref41}
T.~G. Dietterich, ``Ensemble methods in machine learning,'' in \emph{International workshop on multiple classifier systems}.\hskip 1em plus 0.5em minus 0.4em\relax Springer, 2000, pp. 1--15.

\bibitem{ref32}
N.~V. Chawla, K.~W. Bowyer, L.~O. Hall, and W.~P. Kegelmeyer, ``Smote: synthetic minority over-sampling technique,'' \emph{Journal of artificial intelligence research}, vol.~16, pp. 321--357, 2002.

\bibitem{ref33}
T.~Saito and M.~Rehmsmeier, ``The precision-recall plot is more informative than the roc plot when evaluating binary classifiers on imbalanced datasets,'' \emph{PloS one}, vol.~10, no.~3, p. e0118432, 2015.

\bibitem{ref21}
L.~Bellier, A.~Llorens, D.~Marciano, A.~Gunduz, G.~Schalk, P.~Brunner, and R.~T. Knight, ``Music can be reconstructed from human auditory cortex activity using nonlinear decoding models,'' \emph{PLoS biology}, vol.~21, no.~8, p. e3002176, 2023.

\bibitem{ref22}
J.~Berezutskaya, M.~J. Vansteensel, E.~J. Aarnoutse, Z.~V. Freudenburg, G.~Piantoni, M.~P. Branco, and N.~F. Ramsey, ``Open multimodal ieeg-fmri dataset from naturalistic stimulation with a short audiovisual film,'' \emph{Scientific Data}, vol.~9, no.~1, p.~91, 2022.

\bibitem{ref23}
S.~M. Peterson, S.~H. Singh, B.~Dichter, M.~Scheid, R.~P. Rao, and B.~W. Brunton, ``Ajile12: Long-term naturalistic human intracranial neural recordings and pose,'' \emph{Scientific data}, vol.~9, no.~1, p. 184, 2022.

\bibitem{ref42}
R.~J. Zatorre and P.~Belin, ``Spectral and temporal processing in human auditory cortex,'' \emph{Cerebral cortex}, vol.~11, no.~10, pp. 946--953, 2001.

\bibitem{ref24}
M.~O. Memar, N.~Ziaei, B.~Nazari, and A.~Yousefi, ``Rise-ieeg: Robust to inter-subject electrodes implantation variability ieeg classifier,'' \emph{arXiv preprint arXiv:2408.14477}, 2024.

\bibitem{ref43}
L.~Banker and P.~Tadi, ``Neuroanatomy, precentral gyrus,'' 2019.

\bibitem{ref25}
S.~M. Peterson, Z.~Steine-Hanson, N.~Davis, R.~P. Rao, and B.~W. Brunton, ``Generalized neural decoders for transfer learning across participants and recording modalities,'' \emph{Journal of Neural Engineering}, vol.~18, no.~2, p. 026014, 2021.

\end{thebibliography}

\end{document}